*Chapter*

# ROLE OF RADON AS A PRECURSOR TO EARTHQUAKES: AN APPRAISAL

*Rajib Biswas**

Department of Physics, Tezpur University, Tezpur, Assam, Inda

## ABSTRACT

Radon monitoring has recently gained extensive attention among seismologists. It is now widely reported as a precursssory signal prior to occurrence of a seismic event. The enhanced estimation of radon in soil gas is basically attributed to the strain changes which arise within the earth surface. In this chapter, the role of radon as a precurssor is eleborated, acompanied by the anomalies which come across during observation. Additionally, the recent studies done in Himalayan Belt is also reviewd along with future perspectives in radon estimation.

**Keywords**: Radon, Precursor, Emanation, Earthquake

## INTRODUCTION

Literraly, precurssor stands for something that happened or existed before another thing, especially if it either developed into it or had an influence on it. Going by this definition, we refer seismic precursor to trends which in general precede the occurrence of an seismic event. They can be treated as the forerunner of earthquakes. These precurssors comprise of different physiological changes such as deformation or fracture in ground, rise in sea water level, changes in oceanic tides, foreshocks, anomalies in b-values, radon content, water table fluctuations, magneto-telluric currents and many more. Rooted in these observations are the basis of earthquake prediction phenomena. They all contribute to the estmation of fundamental paarmeters of eartqukes which include origin time and magnitude of the impending seismic event. Although large in number, these precurssors suffer from a large drawback. The main problem lies with the effective separation of signals from inherent noise. The most important problem with all these precursors is to distinguish signals from noise. We can not gain any useful information just on the basis of a single precurssor. It has to be a integral approach engaging multitude of presurssors, albeit under the efficient control of an efficent prediction algorithom. Above all, proper evaluation of precurssory phenomena direcly necessitate better comprehension of the physical processes which cause these precursssors. In order to describe these precurssors, one has to resort ot physical model of varied attributes. In general, based on these attributes, the models are grouped in tow classes. The firs one deals with prediction of fault slip behaviour without detaling any variations in material properties surrounding the fault. On the contrary, the second one dewlls upon relations based on bulk, thereby predicting changes in material properties encircling the fault. In such cases, the first type includes nucleation and lithospheric loading models which need special mention. Similarly, the prominent one falling in the secod category is the dilatency model.

---

* Corresponding Author address
　Email:rajib@tezu.ernet.in, rajivb27@gmail.com

Among the aforementioned precurssors, radon estimation has been widely considered as one of the vital parameters. There has been several documentations of monitoring anomalies in emissions of gases like radon, helium, hydrogen, $CO_2$ etc. However, among researchers, radon emerges to be the favourite one; thanks to easy detection procedure. In this chapter, the observation and analysis of radon monitoring will be highlighted. It will also be endeavored to give a outline of the difficuties in the interpretation of data and the relevant foreplay of parameters accompanying anomalous radon emanations. Further, the recent studies across the Himalayan Belt are also reviewd along with future perspectives.

## FUNDAMENTALS OF RADON EMANATION

In genral, Radon occus in three isotope forms. As given in Table 1, Radon 222 having the higest half life is preferred over others. This basically arises as a decay product of 238 U with a uniform distribution over Earths' crust. Being member of 236Ra, it can be easily carried by gases from deeper lyers of earth to the surface.

| Class | Name | Decay Product | Half Life |
| --- | --- | --- | --- |
| Isotope I | Radon 222 | 238U | 3.8 Days |
| Isotope II | Radon 220 | 232Th | 54.5 s |
| Isotope III | Radon 218 | 235U | 3.92 s |

Table 1: Radon Isotopes with their half life

Major part od radiation is basically attributed to the elements such as 218Po letting short lived $\alpha$ particles , accompanied $\beta$ emmision from another 214Bi. Being endowed with inertness, chemical activity of Radon with other elemnts is very poor. As such, it is mostly rooted in the land surface which makes it a valuable precurrsor or as signature for locating active faults [1-2]. When there is an impending large seismic event, there emerges to be enhanced anomalous emmissions of radon which is further assisted by augmented seismic activity in seismic prone areas.

Prior to the ouccurence of a seismic event, there erupts phenomena like dislocations in Rocks, generation and migration of charged partilces, resistivity changes, emission of radon and other gases, mainly caused by microfracturing of Rocks. Precisely, strain fluctuations happening at the earts surface during an earthquake are likely to increase the emanation of radon concentration in soil gas. As reported by Zoran 2002 and Ghosh et al. 2011 [3-4], there had been anamolaous radon emanations, caused by variations in changes preceding medium to large earthquakes.

As for illustrations, the probale radon flux has been pictorially represented with respect to different types of faulting. It is evident that Figs. 1(a) to 1(d) depict several faulting mechanism. Due to movemennt in rock masees, precisely the footwall and hanging wall, microfrcaturing pops up preceding a seismic event. Consequently, we can have several eruptions as mentioned eariler. When there is accumulation of strain, the permeability of rocks varies. This has direct impact in the opening or closing of conduits/passages through which radon can be effectively trasported by carrier gases like $CO_2$, $CH_4$ etc.Keeping aside these carrier gases, we lokk for radon due to its unique properties of having minimized inluences of unrelevant phenomena. Noteworthy point is that the radon emanations possess their individual characteristics as clear from the shape of the radon flux associated with varied faulting. This also establishes itself as a tracer for loation of fauting.

## METHODS OF MEASUREMENT

There are several ways of measuring radon estimations, for example, one can resort to continous mode or discerte mode; albeit, duration of measurement palys a very vital riole.The entry of rado gases into detector chiefly haapens by process of natural diffusion or active technique which need electrical pumping. The widely used detectors are given below.

a. **Solid State nuclear track detectors:** This is most widely used detector among others. They basically comprise of Cr-39 type or LR-115 one. The sensitivity is mostly with rescpec to $\alpha$ particles. Subject to chemical etching, the tracks can be visible under optical microscope. This detector is inexpensive, require no external power. Being sensitive only to $\alpha$ particles, there is little inluence of other external factors such as humidity, temperature variations, light etc.

b. **Electret detector**: This kind of detector comprises of an electrets which exhibits a variation in charge due to its dielectric property. As particle from Radon decay enter the system, there emerges a total change in charge concentration. It has got merits like capability of large duarion storage of record with negligible impact from moisture in the envelope. It is alos easily readable. However, it is crippled by the lacoonae that very low as well as extreme doses are not encompassed by the response curves. Again, it is also susceptible to gamma radiations.

c. **Scintillation detector**: Coupled generally with a photomultiplier tube, it counts the photons which arise due to interactions of $\alpha$ particles as a result of decay of radon. The commercially available type bacially comprises of zinc sulphide scintillation cell

d. **Thermoluminescent detector**: This device is basically is based on thermoluminescence. Maintaing a short separation between themoluminescent material and a metallic plate, one can navigate the electrical charge towards collection. Due to deposition of radon daughters, one can easily estimate energy storage, after proper expositions.

e. **Activated charcoal**: The mechanism of detection is entirely credited to the absorption capability of radon by charcoal. Although simplistic in operational procdure, they are affected by humidity fluctuations. Moreover, they can detect activity only upto couple of days.

There is another class of detector which has now become widely popular among the researchers. With the advent of silicon detectors, the radon monitoring has become highly efficient. Although most of these detectors with a provision of being installed at a certain depth or buried inside the soil, require power; that can be overcome by the recent surge in solar power panel which allow their installation at active fault regions. They are more advantageous as compared to the aforementioned counterparts; thanks to their ability of continous radon measurements with choices in monitoring to be executed in time-dependent or space dependent fluctuations in concentrations.

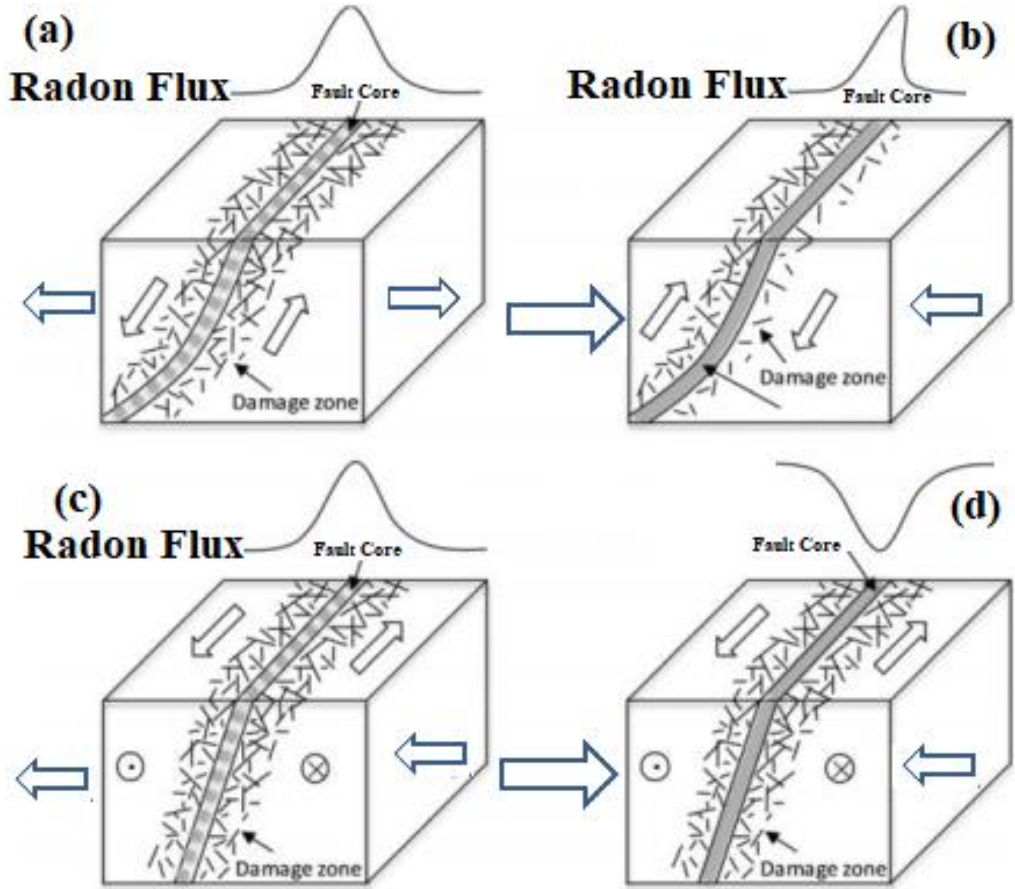

Fig. 1. Relationship between stress regime and Radon flux with respect to (a) Normal Faulting (b) Thrust Faulting (c) Extensional Faulting (d) Compressional Faulting. (Courtesy : Emily Daze: Earthuake Presentations)

## PRINCIPLE OF RADON ANOMALIES

The radon research trend in most cases revolves with report of anaomalies prior to occurrence of seismic event. The randon anomaly is basically defined as execedence of mean radon levele by twice the standard deviation. In this direction, plentiful demonstrations in lab scal as well as in-situ assessment were executed, followed several phenomenological models. The observed anomalies can be ascribed to two basic categories.

a) Production in depth origin

b) Production in local and conveyed by interstitial fluids; mobility of which is triggerd by other geodynamical events.

Between these two, the latter one is preferred because of substantial prrof by experimentalists. As per the dilatancy-diffusion model, the crack growth in the dilatancy volume or fluctualtion in ground water flow could plaisbly trigger radon anomalies. This will lead to closure or expansion of old cracks. In the same way, new openings may also creep in, further accompanied by redistribution of new or old cracks [5-6] (*Scholz,1973; Planinic et al, 2001*). Depending on the type of rocks, this closure or expansion give rise to various transport mechanism for radon. As for instance, there will be prominent effect in the diffusivity of radon in dry rocks owing to opening or closing. In the same direction, the subsurface gas flow may enhance due to volumetric alteration thereby augmenting radon transport. Meanwhile, presence of water in newer cracks gives rise to enhanced transfer of radon from rock matrix to water. In case closure occurs in cracks filled with water, then emanation from rock to water will vary due to migration of water as a result of compression. These happenings contribute overall to the pressure and water level variations in the related aquifier. Although the aforementioned process seems reliable, however, requisite strain or stress has to be considered in far field approximation from the epicenter.

Another theory as an alternative to the above mentioned mechanisms refers to tye corrosion caused by strees.As promalagated by A*nderson* and Grew (1977) [7], the radon anomalies are conjectured to be attributed to the gradual growth in crack incurre by strees corrosion.This in general happen prior to mechanical cracking basically in wet environment. However, radon emanation is totally dependent on factors such as type of rocks, porosity, elasticity as well as hygroscopic properties. Nothwitstading these local paramteres, there is a need of justification for proving that small changes lead to the radon anomalies. This surmise encompasses basically near fault zones. However, the seismically active zomes does not come under this.

In the similar note, anpther plausible theory is rooted in crustal compression. This was reprted by King (1978)[8]. As per his formulation, anomalies in concentration of radon arise due to increase in compressional force, leading to an earthquake. As a consequence, soil-gas is squeezed out to the atmosphere in an enhanced rate. In general, these anaomalies are reported to be observed in at longer distances from the epicenter. This far field emission may be attributed to the changes occurring in the immediate neighbourhood of the radon monitoring station. However, this will be possible provided the strain or stress propagate from source zone to monitioring station. As a result of propagation, there are changes like porosity, flow rate of local ground water in the vicinity of the receiver site. As per estimate, radon detection capability as determined by diffusivity lies basically within a few meters in case of normal porosity and moisture content. The upward surge of carrier with standard displacement of microns per sec gradually augments the concentration of radon near the receiver site which is further subject to depletion of radon at ground level. Keeping in mind the average half life of radon, it is obvious that transportation of radon from lower parts due to decay should have been loner than observed data. In other words, this involves a longer radon column than few meters. This has other intricacies like possible decay of radon before reaching receiver site. In the same note, movile pore fluids have ilnfluences like increasing temporal emanation and simultanoesly exhausting the probable sources of radon.

In addition to these, the radon anomalies are also observed to be influences various types of rocks. As for instance, there is reported to be existence of negative anomalies in case of highly porous rocks as per Tuccimei et al. (2010) [9]. According to them, there is a reduction in pathways due to collapse of core resuting from strain. Even, the radon concentration can vary around the margins of fault system. Acoording to [10], signal grows weaker near the margin of the fault. On the contrary, if the falut system is constituted by crytalling igneous rocks, then there would be negligible detection of radon signal prior to an earthquake. Thanks to the crystalinity which prevents conduits from shaping up. Meanwhile, owing to opening upconduits during a seismic event may give rise to sharp positive anomalies as observed by [11]. According to [12], there is possibility of both –ve and +ve anomalies for identical rock mass; albeit controlled by deformation mechanism. In case of normal faulting, there emerges +ve type where as –ve type erupts in case of cmpressional regimes. In both the cases, the different patter in anaomalous radon concentration could be attributed to the closure or opening of conduits. Although several documentations exist in literarure, there are few which can relate the basic mechanism of anaomalous radon emanation.

## STUIDES ACROSS HIMALAYAN BELT

Himalayn region is known for its seismicity. Many basic as well as observation studies had been perfomed in the last two decades. It has become a customary practice of monitoring seismic activities in Himalayn region as it is believed to be accumulation of stresses which may trigger any seismic event of greater magnitude. This necessatitates a detailed study on active as well as dormant faults in and around Himalayan belt. Ghosh et al., [2007] [13] reported radon anaomaly as a recurssory signal to earthquake. Though a simple Tracketch technique, they monitored radon concentration as measured by solid state detector. However, they concluded the anomalies to be caused by events other seismic evens. In a similar note, Walia et al., [2008] [14] conducted study of anomalous radon as well as helium nearby faults ofDharamsala area, NW Himalayas. They correlated anomalous pattern to different tectonic features and lithographic scales. In another study, Cinti et al., [2009] [15] carried out an extensive survey, engaging thermal water resources of Himachal Pradesh. They showed lot of relevant geochemical parameters by collecting smaples across these sources. As per investigations of Singh et al., [2010] [16] which was conducted in Chamber Valley of NW Himalaya, there exsists strong correlation between radon anaomalies with seismic events in the range of 2.2 to 5.0. Afterwards, Mahajan et al., [2010] [17] tried to establish relationship between anaomalous radon emanation with neotectonic features of NW Himalaya. Apart from these, vital information of tectonic development could also be obtained from anaomalous radon and hellium data as per Mahajan et al. [2010]. Singh et al. 2017[18] reports the analysis of soil radon data recorded in the seismic zone-V, located in the northeastern part of India. They did uninterrupted measurements of soil-gas emission along Chite fault in Mizoram (India). They also carried out investigations in depth domain by incorporating Indo-Burma seismic activity. In another study by Sahoo et al., [19] the soil radon (Rn222) and thoron (Rn220) concentrations were recorded at Badargadh and Desalpar observatories in the Kutch region of Gujarat, India. Their aim was to investigate origin of radon emanation as a recurssory signal to seismic event which was further assisted by insights on the influence of meteorological parameters on radon emission. Hazra et al. repoted abrupt increase in greenhouse gases (like CO2, CH4, H2 etc.) and enhancement of radon emanations. In a recent article by Chowdhury et al. 2019 [20], it is shown that meteorological parameters largely influences radon exhalation rate. The study relates to highly faulted Chhotanagpur Plateau of eastern India.

## FUTURE PERSPECTIVES

Earthquake predictions have so far come a long way. It is basically rooted in systematic observation and analysis of phenomena related to precursory signals. Nothwithstanding growing interest in this sector, there is dearth of concrete theory which can efficiently describe mechanism of earthquakes with reference to precursors. The lacking can be attributed to the complex nature of the interplaying factors and conditions ruling these natural calamities. As a result, the practical difficulties in regards to the prediction methods can not be entirely ruled out until and unless there is in depth understanding of these. It will not be an exaggerartion to assert that rigorous prediction of earthquakes is near to impossibility which is solely based on radon monitoring. However, the general consensus among the research community is that there is variation of radon activity which may precede a earthquake or during an earthquake. It can also be treated as a postearthquake activity. The kind of influences from meterological parameters also hinders the full adoption of radon as a precursor because they in most cases lead to false activity other than seismic events. Untill and unless there is concrtet in depth understanding as well as a concrete synergistic formulation with geophysical parameters, the full potential radon as a earthquake precursor can not be effectively exploited.

## CONCLUSION

In summary, the basics of radon observation are elaborated. The detection procedures are alos detailed. Along with that, the current scenario in relation to the study of radon estimation across the Himalayan Belt is also briefly highlighted. Although there are several reports implying anomalies of radon emanation as a precursory signature, however, the conlusivenes can not be accounted for as most of them are based on correlations and as well as empirical vaues. In such cases, in depth understanding of the physical mechanism is inevitable which will pave way for more foucused research in this direction. On the same note, the data collection has to be for a long duration so as to cover more events in longer span of time, thus ruling out averaging spiking of radon emanation arising from attributes other than seismic events which is genrally observed in small duration of monitoring.

Conflict of Interest: Author RB declares no conflict of interest.

Role of radon as a precursor to earthquakes: An appraisal    13

## REFERENCES

Weinstein, Joshua I. 2009. "The Market in Plato's *Republic*." *Classical Philology* 104:439–58.

[1] Zoran, M. 1979. "Anomalous high concentrations of radon and its alpha active descendents in the lower atmosphere afterward strong Romanian earthquake from March 4, 1977." In: Cornea C, Radu I (eds.) *Seismological researches for Romanian earthquakes from March 4, 1977*, 447–452.

[2] Zoran, M., Savastru, R., Savastru, D., Chitaru, C., Baschir, L., Tautan, M. 2012. "Monitoring of radon anomalies in South-Eastern part of Romania for earthquake surveillance." *J Radioanal Nucl Chem* 293(3):769–781.

[3] Zoran, M. 2002. "Radon in soil variations for Vrancea seismic area." *Revue Roumanie de Geophysique* 46:111–118.

[4] Ghosh, D., Deb, A., Sahoo, S.R., Haldar, S., Sengupta, R. 2011. "Radon as seismic precursor: new data with well water of Jalpaiguri, India." *Natural Haz* 58(3):877–889.

[5] Scholz, C.H., Sykes, L.R. and Aggarwal, YP.1973. "Earthquake prediction: a physical basis." *Science* 181:803-810.

[6] Planinic, J., Radolic, V., Lazanin, Z. 2001. "Temporal variation of radon in soil related to earthquakes." *Applied Radiation and Isotopes* 55:267-272.

[7] Anderson, O.L., and Grew, P.C. 1977. "Stress corrosion theory of crack propagation with application to geophysics." *Rev. Geophys. Space Phys* 15:77-104.

[8] King, C.Y. 1978. "Radon emanation on San Andreas Fault". *Nature* 271: 576-519.

[9] Tuccimei, P., S. Mollo, S. Vinciguerra, M. Castelluccio, and M. Soligo 2010. "Radon and thoron emission from lithophysae-rich tuff under increasing deformation: An experimental study." *Geophys. Res. Lett.* 37:L05305.

[10] Sun, X., Yang, P., Xiang, Y., Si, X., & Liu, D. 2018. Across-fault distributions of radon concentrations in soil gas for different tectonic environments. *Geosciences Journal* 22(2): 227–239.

[11] Mollo, S., Tuccimei, P., Heap, M. J., Vinciguerra, S., Soligo, M., Castelluccio, M., Scarlato, P., and Dingwell, D. B. 2011. "Increase in radon emission due to rock failure: An experimental study." *Geophys. Res. Lett.* 38: L14304.

[12] Tuccimei, P., Mollo, S., Soligo, M, Scarlato, P., Castelluccio, M. (2015), "Real-time setup to measure radon emission during rock deformation: implications for geochemical surveillance." *Geosci. Instrum. Method. Data Syst* 4(1): 111-119.

[13] Ghosh, D., Deb, A., Sengupta, R., Patra, K.K., Bera, S. 2007. "Pronounced soil radon anomaly- Precursor of recent earthquake in India." *Radiation measurement* 42: 466-471.

[14] Walia V., Mahajan S., Kumar A., Singh S., Bajwa B.S., Dhar S., Yang T. F. 2008. "Fault delineation study using soil-gas method in Dharamshala area, NW Himalayas, India." *Radiat. Meas.* 43: S337-S342.

[15] Cinti, D., Pizzine L., Voltattorni N., Quattrocchi F., Walia V., (2009) "Geochmistry of thermal fluids along faults segments in the Beas and Parvati alley (north-west Himalaya Himachal Pradesh) and Sohana town (Haryana), India." *Geochemical Journal* 43:65-76.

[16] Singh, S., Kumar, A., Bajwa, B.S., Mahajan, S., Kumar, V., Dhar, S. 2010. "Radon Monitoring in Soil Gas and Ground Water for Earthquake Prediction study in NW Himalaya, India." *Terr. Atmos. Ocean Sci* 21(4 ): 685-695.

[17] Mahajan, S., Walia, V., Bajwa, B. S., Kumar, A., Singh, S., Seth, N., Dhar, S., Gill, G. S. and Yang, T. F. 2010. "Soil-gas radon/helium surveys in some neotectonic areas of NW Himalayan foothills, India." *Natural Hazards and Earth System Sciences* 10(6):1221—1227.

[18] Singh, S., Jaishi, H.P., Tiwari, R.P. et al. 2017. "Time Series Analysis of Soil Radon Data Using Multiple Linear Regression and Artificial Neural Network in Seismic Precursory Studies." *Pure Appl. Geophys* 174: 2793.

[19] Sahoo, S.K., Katlamudi, M., Shaji, J.P. et al. 2018. "Influence of meteorological parameters on the soil radon (Rn222) emanation in Kutch, Gujarat, India." *Environ Monit Assess* (2018) 190:111.

[20] Chowdhury, S., Barman, C., Deb, A. et al. 2019. "Study of variation of soil radon exhalation rate with meteorological parameters in Bakreswar–Tantloi geothermal region of West Bengal and Jharkhand, India." J *Radioanal Nucl Chem* (2019) 319: 23.